\documentclass[a4paper,11pt]{article}
\usepackage{aaskaiid}
\usepackage{orcidlink}

\title{Synergies Between Pulsar Timing Array and Astrometry}
\ShortTitle{Synergies between pulsar timing array and astrometry}

\author[1,2,3]{Gabriele Perna\orcidlink{0000-0002-7364-1904}}
\ShortName{Perna et al.}
\emailAdd{gabriele.perna@phd.unipd.it}
\author[1,2,4]{Nicola Bellomo\orcidlink{0000-0002-4375-705X},}
\emailAdd{nicola.bellomo@unipd.it}
\author[1,2,4]{Daniele Bertacca\orcidlink{0000-0002-2490-7139}}
\emailAdd{daniele.bertacca@unipd.it}
\affiliation[1]{Dipartimento di Fisica e Astronomia ``Galileo Galilei" Universit\`a di Padova, Via Marzolo 8, I-35131 Padova, Italy}
\affiliation[2]{Istituto Nazionale di Fisica Nucleare (INFN), Sezione di Padova, Via Marzolo 8, I-35131 Padova, Italy}
\affiliation[3]{National Institute for Chemical Physics and Biophysics (NICPB), Ravala 10, 10143 Tallinn, Estonia}
\affiliation[4]{Istituto Nazionale di Astrofisica (INAF), Osservatorio Astronomico di Padova, vicolo dell’Osservatorio 5, I-35122 Padova, Italy}

\abstract{
The presence of a gravitational wave background can be established not only via exquisitely precise pulsar timing array (PTA) measurements, but also via astrometric observations.
In fact, the very same background responsible for the delay in the arrival time of pulse is also responsible of an apparent displacement of galactic objects as stars and asteroids.
In this chapter we explore the natural synergy between the SKA Observatory, and current/future astrometric probes of the position of Milky Way objects.
On top of presenting the potential of SKAO alone in terms of detecting a gravitational wave background, we also demonstrate the increased sensitivity that is actually achievable when SKAO measurements are used in combination with astrometric ones.
In particular, we observe an approximate improvement ranging from~$10\%$ up to~$50\%$ in terms of forecast sensitivity for a PTA-astrometry joint-analysis.
}

\begin{document}
\maketitle

\section{Introduction}

Thanks to observational advances, in the recent years GW detections have revolutionized our understanding of the Universe. 
A specific breakthrough point for the community has been 
the reported evidence of a gravitational-wave background (GWB) in the nHz frequency band by Pulsar Timing Array (PTA) experiments~\citep{NANOGrav:2023gor, EPTA:2023fyk, Reardon:2023gzh, Xu:2023wog, Miles_2024}.
Despite this experimental success, one of the major challenges the community will face is establishing if the origin of this GWB is astrophysical or cosmological~\citep{Agazie_2023_smbhb, Afzal_2023}.
Irrespective of the frequency band of interest, the detection of an astrophysical GWB would provide information on the population of astrophysical GW sources too weak or too abundant to be individually resolved~\citep{sesana_stochastic_2008, Ferrari:1998ut, Ignatiev:2001jr, 2004MNRAS.351.1237H, Regimbau:2011rp}. 
If confirmed, it could shed light on different aspects of the physics of compact objects, as their mass spectrum, spin, formation mechanism, and connection with their astrophysical hosts~\citep{Regimbau:2011rp, chen_constraining_2019}. 
On the other hand, in the case of a cosmological origin, several early-Universe scenarios predict a background of GW generated by phenomena such as inflation, cosmic strings, or first-order phase transitions~\citep{Bartolo:2016ami, Bartolo:2019yeu, Guzzetti:2016mkm, Caprini:2018mtu, Maggiore:2007ulw}. 
Therefore, its detection would represent a unique opportunity to glimpse into the primordial Universe and the physical processes that shaped those epochs. 

Given the challenges ahead, a series of third generation GW detectors will come online in the next decade, with the goal of covering as much as possible the GW frequency band, in an attempt to close observational gaps between the different frequency bands that currently limit our access to the entire GW spectrum.
In this context, by the time the SKAO is fully operational, several other detectors are also expected to start taking data, as LISA~\citep{LISA:2017pwj} in the mHz band, and Einstein Telescope~\citep{Sathyaprakash:2011bh, LIGOScientific:2016wof, Maggiore:2019uih, ET:2025xjr} and Cosmic Explorer~\citep{Reitze:2019iox} in the Hz-kHz band. 
On the other hand, in the nHz band, the SKAO will carry forward the success story of PTA experiments~\citep{Janssen:2014dka} by probing an unprecedented number of millisecond pulsars across the southern hemisphere (see also \cite{Keane01.2026.SKA,Shannon01.2026.SKA}). 

In this Chapter we focus on potential synergies between pulsar timing and astrometric measurements, which has the potential of expanding even further the observed GW frequency range, see, e.g.,~\cite{Wang:2022sxn} for the case of astrometry with photometric surveys. 
The idea of combining apparently disconnected experiments is motivated by the fact that any GW passing through the propagation path of photons induces a deviation in its trajectory that results in observing a small fluctuation not only in the photons time-of-arrival, but also in their observed angular position of emission, which will be detectable with high-precision astrometry measures~\citep{Book:2010pf, Braginsky:1989pv, Damour:1998jm, Pyne:1995iy, Jaffe:2004it, Kaiser:1996wk, Linder:1986fdo}.

The first detailed characterization of the expected astrometric deflection signal induced by a GWB was provided by~\cite{Book:2010pf} in the ``distant source'' limit, i.e., under the assumption that the distance from the observer to the source emitting photons is significantly larger than the GW wavelength. 
This approximation is well motivated in the context observations of distant quasars~\citep{Pyne:1995iy}, and it allowed to place one of the first upper bounds on the dimensionless GW energy density $\Omega_{\mathrm{GW}}$~\citep{Gwinn:1996gv, Titov_2011, Darling:2018hmc}.
This technique has then be refined by~\cite{Aoyama:2021xhj, Wang:2022sxn, Jaraba:2023djs} in the context of quasars, and by~\cite{Mentasti:2023gmr} for sources orbiting in the Solar System. 
Recently,~\cite{Perna:2025bdu} developed for the first time an approach that does not require the assumption of the ``distant source'' limit, which we will employ in this Chapter. 

The potential of astrometric measurements is outstanding: since the only intrinsic limitations are represented by the cadence and total time of the observations, they can easily fill the existing gap between PTA and LISA characteristic frequencies~\citep{Caliskan:2023cqm}. 
Additionally, future facilities with high observing cadence, such as the~\textit{Nancy Grace Roman} Space Telescope~\citep{nasa_roman}, may further extend the probed frequency range beyond~$10^{-4}$ Hz (see also~\cite{Rioja01.2026.SKA} for predictions on ultra-precise astrometry using the SKA). 
The scientific potential is also unprecedented.
The intrinsic vector nature of astrometric deflections allows the construction of parity-odd correlation functions, and cross-correlations between astrometry and PTA data can provide a powerful probe of possible chiral components of the GWB~\citep{Qin:2018yhy}. 
Partial overlap in frequency range, combined with the distinct signatures that a common GWB imprint on PTA and astrometric observables, highlight the importance of cross-correlating the two datasets to enhance the precision of both observational methods~\citep{Caliskan:2023cqm, Vaglio:2025tex}. 
In summary, although the SKAO will represent a true scientific milestone on its own, the full potential of this experiment will be unlocked only when its observations are combined with complementary astronomical ones. 
In light of this fact, in this Chapter we assess the potential scientific gain in adopting a synergistic approach between SKAO and astrometric observations to detect and characterize the presence (and properties) of a GWB.

This short chapter is organized as follows.
In Sec.~\ref{sec:angular_deflection} we briefly review the theory underlying angular position deflections generated by an intervening GWB.
In Sec.~\ref{sec:pta_astrometry_analysis} we report the main steps of the analysis, including the characterization of the signal in harmonic space, a brief overview of the different astrometric tracers, and the forecast of the improvement in sensitivity due to this synergistic approach. 
Finally, we conclude in Sec.~\ref{sec:conclusions}.


\section{Angular position deflection induced by a GWB}
\label{sec:angular_deflection}

In this section, we briefly review the theory behind the angular deflection sourced by a passing GW.
The photon unperturbed world-line can be defined as
\begin{equation}
    x^\mu (\lambda) = (t_0+\omega_0 \lambda, -\omega_0 \lambda \hat{\mathbf{n}}),
\end{equation}
where~$t_0$ and~$\mathbf{x}=\mathbf{0}$ are the time and position where the observation is performed, $\omega_0$ is the photon unperturbed frequency, $\lambda$ is an affine parameter, and~$r_s$ is the distance of the photon source in the direction~$\hat{\mathbf{n}}$. 
In this framework, the angular deflection on the positions of an astrophysical object due to a intervening, monochromatic, GW is given by~\citep{Book:2010pf}
\begin{equation}
    \delta n^i = \left(\delta^K_{ik} - n^i n^k\right) n^j \left[\frac{1}{2} h_{jk}(\lambda_0) - \left( \delta^K_{kl} - \frac{p^k n^l}{2(1+\hat{\mathbf{p}}\cdot\hat{\mathbf{n}})} \right) \left( h_{jl}(\lambda_0) - \frac{1}{\lambda_s}\int_0^{\lambda_s} \!\!\!\!\! d\lambda\ h_{jl}(\lambda) \right)\right],
\end{equation}
where the GW direction and frequency are labeled as~$\hat{\mathbf{p}}$ and~$f$, respectively. 
More explicitly, we have~\citep{Book:2010pf}
\begin{align} \label{dn2}
     \delta n^{{i}} =  &  \Bigg( \Bigg\{ 1 + \frac{i(2 + \hat{\mathbf{p}}\cdot \hat{\mathbf{n}})}{\omega _0 \lambda _s \Omega (1+\hat{\mathbf{p}}\cdot \hat{\mathbf{n}})}[1 - e^{-i\Omega \omega _0 (1+\hat{\mathbf{p}}\cdot \hat{\mathbf{n}})\lambda _s}]\Bigg\} n^i \nonumber \\
     &\quad + \Bigg\{1+ \frac{i}{\omega _0 \lambda _s \Omega (1+\hat{\mathbf{p}}\cdot \hat{\mathbf{n}})}[1 - e^{-i\Omega \omega _0 (1+\hat{\mathbf{p}}\cdot \hat{\mathbf{n}})\lambda _s}]\Bigg\} p^i \Bigg) \frac{n^j n^k \mathcal{H}_{jk}e^{-i\Omega t_0}}{2(1+\hat{\mathbf{p}}\cdot \hat{\mathbf{n}})} \nonumber \\
     &\quad - \Bigg\{ \frac{1}{2} + \frac{i}{\omega _0 \lambda _s \Omega (1+\hat{\mathbf{p}}\cdot \hat{\mathbf{n}})}[1 - e^{-i\Omega \omega _0 (1+\hat{\mathbf{p}}\cdot \hat{\mathbf{n}})\lambda _s}]\Bigg\} n^j \mathcal{H}^i _j e^{-i \Omega t_0} \,, 
\end{align}
where~$\Omega=2\pi f$, $\mathcal{H}_{ij}$ contains information regarding the amplitude and polarization of the GW, and the affine parameter~$\lambda_s$ is evaluated at the source location
\begin{equation} \label{lambdas}
    \lambda_s = - \frac{|\mathbf{x}_s|}{\omega_0} = - \frac{r_s}{\omega_0}\,. 
\end{equation}
Since the stochastic deflection created by the background varies over time, the source observed at position~$\mathbf{n} + \delta \mathbf{n}$ will effectively exhibit an apparent proper motion in the sky.

Eq.~\eqref{dn2} can be further simplified in the case of the ``distant source'' limit, i.e. when the distance of the source is greater than the wavelength of the GW causing the deflection. 
Taking the limit~$\Tilde{x} := 2\pi f r_s \to \infty$, the angular deflection can be rewritten as 
\begin{equation} \label{dn.limit}
    \delta n^{i}(t_0, \hat{\mathbf{n}}) =  (n^i + p^i)\frac{n^j n^k \mathcal{H}_{jk}e^{-i\Omega t_0}}{2(1+\hat{\mathbf{p}}\cdot \hat{\mathbf{n}})} - \frac{1}{2}\mathcal{H}_{ij}n_j e^{-i\Omega t_0} \,.
\end{equation}
On the other hand, in the ``short distance'' limit where~$\tilde{x}\ll 1$, \cite{Mentasti:2023gmr} obtained that the angular deflection reads as
\begin{align}
    (\delta n)^i &= \left[ \frac{(1-\hat{\mathbf{p}}\cdot \hat{\mathbf{n}} )n^i n^j n^k \mathcal{H}_{jk}}{2(1+\hat{\mathbf{p}}\cdot \hat{\mathbf{n}} )} + \frac{1}{2}n^j \mathcal{H}^i _j \right] e^{-i2\pi f t_0} + \mathcal{O}(\Tilde{x}).
\end{align}
Although these limiting cases serve as easy-to-understand benchmark results, the combination of different astrometric probes allows to span the entire range of~$\tilde{x}$.
For this reason, in the following we adopt the generic results from~\cite{Perna:2025bdu}, which are based on the general formula in Eq.~\eqref{dn2}, and allow to consistently account for the full 3D distribution of astrophysical sources.


\subsection{Synergies with pulsar timing residual measurements}

Pulsar timing residuals are also affected by the passage of the same GWs that produced the angular deflection of astrometric tracers.
Therefore, it is noteworthy to investigate the potential of combining these two seemingly unrelated probes.
Consider a pulsar at a distance~$\mathbf{r}$ and in the direction~$\hat{\mathbf{n}}$ that periodically emits pulses observed in radio frequency.
The difference in the estimated period between pulses~$\Delta T_\mathrm{GWB}$ due to a GWB, also called ``redshift'', can be written as
\begin{equation}
    z(t,\mathbf{r}) = \frac{\Delta T_\mathrm{GWB}}{T_\mathrm{psr}} = \sum_\lambda \int df d\hat{\mathbf{p}}\ h_\lambda(f,\hat{\mathbf{p}}) \frac{n^i n^j e^\lambda_{ij}(\hat{\mathbf{p}})}{2(1+\hat{\mathbf{n}} \cdot \hat{\mathbf{p}})} e^{-2\pi i ft} \left[1 - e^{2\pi i f r(1 + \hat{\mathbf{p}} \cdot \hat{\mathbf{n}})} \right].
\label{eq:redshift_timing_residuals}
\end{equation}
Given that all the observed pulsars are at distances of order~$\mathcal{O}(\rm{kpc})$, PTA experiments work in the long-arm limit. 
This allows to neglect the last term in the square brackets (usually called pulsar term)~\cite{Roebber:2016jzl} leading to a distance-independent expression
\begin{equation}
    z(t,\hat{\mathbf{n}}) = \sum_A \int df \int d\hat{\mathbf{p}}\ h_A(f,\hat{\mathbf{p}}) \frac{n^i n^j e^A_{ij}(\hat{\mathbf{p}})}{2(1+\hat{\mathbf{n}} \cdot \hat{\mathbf{p}})} e^{-2\pi i ft}\,.
\end{equation}


\section{PTA-Astrometry joint-analysis}
\label{sec:pta_astrometry_analysis}


\subsection{Harmonic analysis}

The redshift is a scalar, thus we will use the well-known expansion in spherical harmonics given by
\begin{equation}
    z(\hat{\mathbf{n}}) = \sum_{\ell m} z_{\ell m} Y_{\ell m}(\hat{\mathbf{n}}),
\end{equation}
where~$Y_{\ell m}$ are the spherical harmonics, and the harmonic coefficients read as 
\begin{equation}
    z_{\ell m} = \int d\hat{\mathbf{n}}\ z(\hat{\mathbf{n}}) Y_{\ell m}^*(\hat{\mathbf{n}}) = \int d\hat{\mathbf{n}} \frac{n_i n_j h_{ij}}{2(1+\mathbf{n}\cdot\hat{\mathbf{p}})} Y_{\ell m}^*(\hat{\mathbf{n}}) \,.
\end{equation}

On the other hand, the angular deflection is a vector; thus, it requires a decomposition in vector spherical harmonics, as, for instance, in the case of E and B modes of the Cosmic Microwave Background.
Therefore, in this case, we have that the angular deflection is given by
\begin{equation} \label{vsh}
    (\delta n)_i = \sum _{l m}\left[ E_{l m}Y^E _{(l m),i}(\hat{\mathbf{n}}) + B_{l m}Y^B _{(l m),i}(\hat{\mathbf{n}})\right] \,,
\end{equation}
where~$Y^E _{(l m),i},\,Y^B _{(l m),i}$ are the components of the vector spherical harmonics defined as~\citep{Varshalovich:1988ifq}
\begin{equation}
    \mathbf{Y}^E_{\ell m} = -\frac{1}{\sqrt{\ell(\ell+1)}} \nabla_\Omega Y_{\ell m}, \qquad \mathbf{Y}^B_{\ell m} = \frac{-i}{\sqrt{\ell(\ell+1)}} (\hat{\mathbf{n}} \times \nabla_\Omega) Y_{\ell m}, 
\end{equation}
with $\nabla_\Omega$ being the angular part of the gradient operator.
Vector spherical harmonic coefficients are generically given by
\begin{align} \label{def_elm}
    X_{l m} & = \int d\hat{\mathbf{n}}\ (\delta n ) ^i Y^X_{(l m),i}(\hat{\mathbf{n}})  \,,
\end{align}
which, in the case of the E- and B-modes of astrometric measurements reduce to
\begin{align}
    E_{\ell m} &= \int d\hat{\mathbf{n}}\ \delta\mathbf{n}\cdot \mathbf{Y}^{E*}_{\ell m} = -\frac{1}{\sqrt{\ell(\ell+1)}}\int d\hat{\mathbf{n}}Y^{*}_{\ell m}(\hat{\mathbf{n}}) (\nabla_\Omega\cdot \delta\mathbf{n}), \\
    B_{\ell m} &= \int d\hat{\mathbf{n}}\ \delta\mathbf{n}\cdot \mathbf{Y}^{B*}_{\ell m} = -\frac{i}{\sqrt{\ell(\ell+1)}}\int d\hat{\mathbf{n}} Y_{\ell m}^*(\hat{\mathbf{n}}) (\hat{\mathbf{n}}\times \nabla_\Omega)\cdot \delta\hat{n}.
\end{align}

In the case of the ``distant source'' limit, we can substitute Eq.~\eqref{dn.limit} and obtain the known form
\begin{align}
    E_{\ell m} &= -\frac{1}{\sqrt{\ell(\ell+1)}}\int d\hat{\mathbf{n}}Y_{\ell m}^*(\hat{\mathbf{n}}) \frac{n_i n_j h_{ij}}{1+\mathbf{n}\cdot\hat{\mathbf{p}}}\nonumber\,,\\
    B_{\ell m} &= -\frac{i}{\sqrt{\ell(\ell+1)}}\int d\hat{\mathbf{n}} Y_{\ell m}^*(\hat{\mathbf{n}}) \varepsilon_{kij} \frac{h_{mj}p^kn^mn^i}{1+\mathbf{n}\cdot\hat{\mathbf{p}}}\,.
\end{align}
These results are reported only to suggest the commonalities between astrometric and pulsar timing measurements in the ``distant source'' limit.
In this sense, in this limit the harmonic coefficients of the E-mode and redshift are related to each other, since
\begin{align}
    z_{\ell m } = - \frac{\sqrt{\ell(\ell+1)}}{2} E_{\ell m}.
\end{align}
This relation immediately suggests that there exists a non-vanishing correlation between pulsar timing and astrometric measurements. 
Since at low multipoles~$z_{\ell m} \sim E_{\ell m}$, we have that in terms of expectation values~$\langle z_{\ell m} z^*_{\ell m}\rangle\sim\langle E_{\ell m} E^*_{\ell m}\rangle\sim\langle E_{\ell m} z^*_{\ell m}\rangle$. 
In other words, at low multipoles, the cross-correlation between~$E$ and~$z$ modes is of the same order of the auto-correlation without being affected by any common source of noise.
Therefore, its inclusion is fundamental not only to improve the constraining power of astrometric observations, but also to cross-check that the measured signal is not contaminated by systematics effects, strengthening the motivation for a joint-analysis of these two complementary probes~\citep{Qin:2018yhy}. 

Irrespective of these considerations, the results of our analysis have been obtained using the fully analytical formalism developed by~\cite{Perna:2025bdu}.
As demonstrated in that work, the auto- and cross-correlation angular power spectra for the redshift and E-mode are
\begin{align}
    C_\ell^{zz} &= \frac{3 H_0^2}{\pi}\left(\frac{(\ell-2)!}{(\ell+2)!}\right)\int_{0}^{+\infty} df  \frac{\Omega_{\rm GW}(f)}{f^3}\,, \\
    C_\ell^{EE}(r,r') &= \frac{12 H_0^2}{\pi} \frac{1}{\ell(\ell+1)}\left(\frac{(\ell-2)!}{(\ell+2)!}\right)\int_{0}^{+\infty} df  \mathrm{Re} \left[ F^E_\ell(f, r) \left[F^E_\ell(f, r') \right]^* \right]\frac{\Omega_{\rm GW}(f)}{f^3}\,,
\end{align}
where~$H_0$ is the Hubble constant and~$\Omega_{\rm GW}(f)$ the dimensionless energy density; thus
\begin{align}
    C_\ell^{Ez}(r) = \frac{6 H_0^2}{\pi} \frac{1}{\sqrt{\ell(\ell+1)}}\left(\frac{(\ell-2)!}{(\ell+2)!}\right)\int_{0}^{+\infty} df\, \mathrm{Re} \left[ F^z_\ell(f) \left[F^E_\ell(f, r) \right]^* \right]  \frac{\Omega_{\rm GW}(f)}{f^3}\,.
\end{align}
On the other hand, the B-mode angular power spectrum reads as
\begin{align}
    C_\ell^{BB}(r,r') = \frac{12 H_0^2}{\pi} \frac{1}{\ell(\ell+1)}\left(\frac{(\ell-2)!}{(\ell+2)!}\right)\int_{0}^{+\infty} df \,\mathrm{Re} \left[ F^B_\ell(f, r) \left[F^B_\ell(f, r') \right]^* \right] \frac{\Omega_{\rm GW}(f)}{f^3}\,.
\end{align}

The $F_\ell^X$ functions appearing for both the E- and B-modes have been introduced by~\cite{Perna:2025bdu} to fully account for the dependence on the distance of the monitored astronomical objects and extend the allowed frequency range.
As expected, in the ``distant source'' limit, we have~$F^X\to 1$.
These expressions also show how both the amplitude and the spectral shape of the GW power spectrum play an important role in determining the amount of correlation between these different probes.

Finally, we emphasize that the standard expectation for any cross-spectrum between B-, and E-modes or redshift is zero because of the different parities of the E- and B-mode signals.
However, the possible existence of chirality in the GWB would induce a non-vanishing cross-correlation, and thus it could be used to test the presence of parity violation not only in the GW signal itself, but also in the physical mechanisms that generated such a signal. 
For instance, in the case of a signal with cosmological origin, a non-zero cross-correlation would indicate the presence of parity-violating phenomena in the primordial Universe, see, e.g., the work of~\cite{Qin:2018yhy, Golat:2022hjf, Liang:2023pbj}.


\subsection{Detectability estimate and Fisher analysis}

We estimate the capability of SKAO to detect a GWB in correlation with astrometric probes by evaluating the cumulative angular signal-to-noise (SNR) ratio for different total observation times and the number of monitored pulsars. 
In full generality, our Signal-to-Noise ratio (SNR) is defined as
\begin{equation}
    \mathrm{SNR}^2 = \sum_{\ell=2}^{\ell_{\rm max}} \mathrm{SNR}^2_{\ell} = \sum_{\ell=2}^{\ell_{\rm max}} \frac{2\ell+1}{2} \mathrm{Tr} \left[\mathcal{C}_\ell \left(\mathcal{C}_\ell+\mathcal{N}_\ell\right)^{-1} \mathcal{C}_\ell \left(\mathcal{C}_\ell+\mathcal{N}_\ell\right)^{-1} \right],
\end{equation}
where the (symmetric) covariance matrix of the signal reads as
\begin{equation}
    \mathcal{C}_\ell = 
    \begin{pmatrix}
    C^{zz}_\ell & C^{zE_1}_\ell & \cdots & C^{zE_n}_\ell & C^{zB_1}_\ell & \cdots & C^{zB_n}_\ell \\
    & C^{E_1 E_1}_\ell & \cdots & C^{E_1 E_n} & C^{E_1 B_1}_\ell & \cdots & C^{E_1 B_n} \\
    & & \ddots & \vdots & \vdots & \ddots & \vdots \\
    & & & C^{E_n E_n} & C^{E_n B_1}_\ell & \cdots & C^{E_n B_n} \\
    & & & & C^{B_1 B_1}_\ell & \cdots & C^{B_1 B_n} \\
    & & & & & \ddots & \vdots \\
    & & & & & & C^{B_n B_n}_\ell \\
    \end{pmatrix},
\end{equation}
and the noise covariance matrix is given by
\begin{equation}
    \mathcal{N}_\ell = \mathrm{diag}\left( N^{zz}_\ell, N^{E_1 E_1}_\ell, \cdots, N^{E_n E_n}_\ell, N^{B_1 B_1}_\ell, \cdots, N^{B_n B_n}_\ell \right).
\end{equation}
The angular power spectra of the noise of redshift and astrometric measurements read as~\citep{Caliskan:2023cqm, Perna:2025bdu}
\begin{equation}
\label{Eq::nois}
    N^{zz}_\ell = \frac{4\pi}{N_\mathrm{psr}} \int df |W_f(f,f_j,\Delta f_j)|^2 S_n(f), \qquad N^{X_{k}X_{k}} = \frac{2\pi \sigma^2_\mathrm{tot}}{N_\mathrm{obj}}\;,
\end{equation}
where~$S_n(f)$ is the one-sided noise power spectrum, $N_\mathrm{obj}$ and $N_\mathrm{psr}$ are the number of astrometric objects and pulsars considered in the analysis, respectively, and~$\sigma^2_\mathrm{tot}$ is the total astrometric error on the angular position.

At the same time, we also adopt a Fisher matrix approach~\citep{fisher:fisher, bunn:fisher, vogeley:fisher, tegmark:fisher} to estimate the expected sensitivity of a future joint-analysis.
In particular, we focus on a GW signal in agreement with the latest results from PTA analysis~\citep{NANOGrav:2023gor}, where
\begin{align}
    \Omega_{\rm GW} = A \left(\frac{f}{f_s}\right)^\gamma\,,
\end{align}
with $A = 2.2\cdot10^{-7}$, $ f_s = 3\cdot10^{-8}$ Hz and $\gamma\simeq 1.63$. 
Since in this case we are mostly interested in estimating the amplitude~$A$ of the GWB, we write our Fisher matrix as
\begin{equation}
    F_{AA} = \sum^{\ell_\mathrm{max}}_{\ell=2} \frac{2\ell+1}{2} \mathrm{Tr}\left[ \left(\partial_{A} \mathcal{C}_\ell \right) \left(\mathcal{C}_\ell+\mathcal{N}_\ell\right)^{-1} \left(\partial_{A} \mathcal{C}_\ell \right) \left(\mathcal{C}_\ell+\mathcal{N}_\ell\right)^{-1} \right]\,,
\end{equation}
and compute the forecast error as~$\sigma_{\rm A} = \sqrt{F^{-1}_{AA}}$.

In terms of astrometric probes, we consider two candidate populations of Milky Way stars and Main Belt asteroids.
The properties of these populations, including their spatial distribution and typical radial and angular position errors, are extensively discussed in~\cite{Perna:2025bdu}.
On the astrometric experimental side, we considered the possible detection of E- and B-modes by a future GAIA-like and Roman-like satellites withe same cadence of observations and total observing time of their predecessors, but with improved angular resolution, as reported in~\cite{Perna:2025bdu}. 
Specifically, similarly to previous works~\citep{Mentasti:2023gmr, Caliskan:2023cqm}, we consider a cadence~$\Delta t =  14\ \mathrm{days}$ and~$\Delta t = 30\,{\rm min}$, and an observation time of~$T= 10\,{\rm years}$ and~$T= 5\,{\rm months}$ for a GAIA-like and Roman-like experiment, respectively.

Focusing on SKAO, we consider different total observation times of 5, 10, and 20 years, and different numbers of monitored pulsars to illustrate their impact on the final SNR. 
The SKAO telescopes are expected to monitor about~$200$ millisecond pulsars with a fortnightly cadence; however, current estimates suggest the possible detection of up to 1000 millisecond pulsars for the design baseline.

\begin{figure}[ht]
    \centerline{
    \includegraphics[width=1\linewidth]{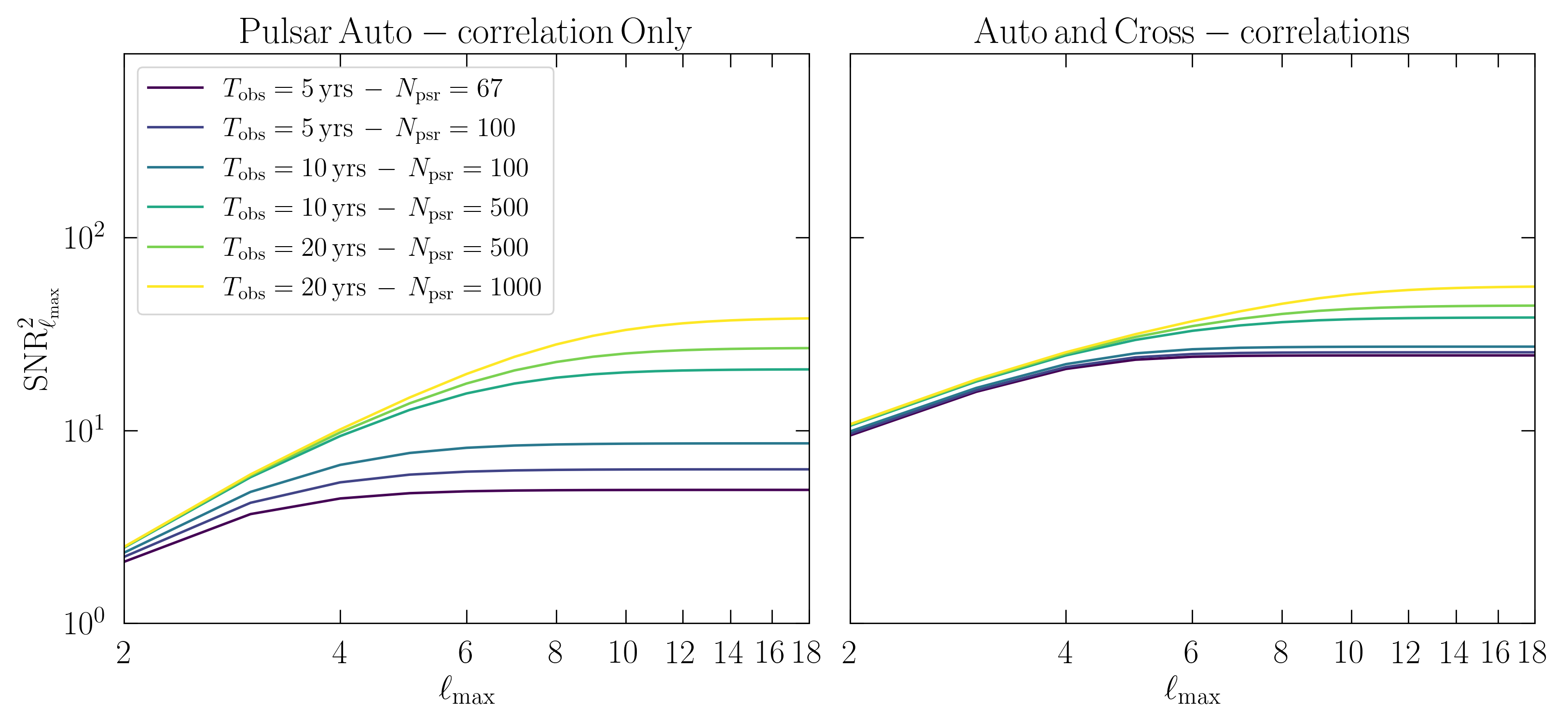}}
    \caption{\textit{Left panel}: Cumulative angular SNR for different observation times and different numbers of monitored pulsars considering only pulsar auto-correlations. 
    \textit{Right panel}: Same as left panel, but with the inclusion of E-modes and B-modes auto-correlations and E-z cross correlations.
    Figure adapted from \cite{Perna:2025bdu}.}
    \label{fig:results}
\end{figure}

We compute the capability of SKAO to detect a GWB by a measurement of the redshift signal alone, and in cross-correlations with additional astrometric probes.
Our results are shown in Fig.~\ref{fig:results}, where we report the cumulative angular SNR obtained considering only pulsar auto-correlation terms (left panel) and including also astrometric E and B mode auto correlations and non-vanishing cross-correlations with the redshift (right panel). 
The figure shows that redshift measurements alone already yield a SNR larger than unity, highlighting the scientific potential of SKAO alone. 
On the other hand, the inclusion of the auto- and cross-correlations of the E- and B-mode astrometric deflections significantly enhances the total angular SNR, going from~${\rm SNR}_{\ell=2}^2\simeq2$ for only pulsars up to ${\rm SNR}_{\ell=2}^2\simeq10$ when also astrometric probes are included. 
This result demonstrates the practical importance of exploiting existing and future synergy between pulsar-timing datasets and astrometric measurements. 

In addition, we study the impact on the SNR of different observation times and monitored pulsars. 
Increasing the observation time improves the sensitivity, effectively reducing the amount of instrumental and timing-model noises the longer a pulsar is observed. 
Similarly, increasing the number of monitored pulsars enhances the detectability of a GW signal. 
We find that already after 5 years, and considering only the pulsars currently monitored by PTA experiments, the multipole $\ell = 2$ can be detected with an SNR greater than 1 (as expected, given the existing evidence for the Hellings–Downs correlation from present-day PTA datasets).
The improved sensitivity of the SKAO relative to current PTA experiments is reflected in the higher SNR obtained when including higher multipoles (up to $\ell = 4$).
We also find that doubling the number of observed pulsars increases the total SNR within the same observing time, making the multipole~$\ell = 5$ relevant for the SNR computation.
Our estimates suggest that, after 20 years of observations, SKAO could detect multipoles up to~$\ell \sim 8$, with~${\rm SNR}^2\simeq44.3$ for five hundred pulsars, and up to~$\ell \sim 10$ with~${\rm SNR}^2\simeq55.7$ in the most idealistic scenario where roughly one thousand pulsars are monitored.
The removal of the astrometric signal from the analysis leads to a decrease in detectability to~${\rm SNR}^2\simeq26.7$ and~${\rm SNR}^2\simeq38.2$, respectively.
We expect the improvement in sensitivity due to a joint-analysis to be crucial, as it enables a more detailed characterization of the GW signal over a wider frequency range, thereby increasing the constraining power of this method.

\begin{table}[ht]
    \centerline{
    \begin{tabular}{lccc}
        \hline
        $T_{\rm obs}$ (yrs) & $N_{\rm psr}$ & Auto-z & Auto+Cross \\
        \hline\hline
        5  &  67   & $9.93\cdot10^{-8}$ & $4.45\cdot10^{-8}$ \\
        5  & 100   & $8.78\cdot10^{-8}$ & $4.37\cdot10^{-8}$ \\
        10 & 100   & $7.52\cdot10^{-8}$ & $4.22\cdot10^{-8}$ \\
        10 & 500   & $4.8\cdot10^{-8}$  & $3.55\cdot10^{-8}$ \\
        20 & 500   & $4.26\cdot10^{-8}$ & $3.3\cdot10^{-8}$ \\
        20 & 1000  & $3.56\cdot10^{-8}$ & $2.95\cdot10^{-8}$ \\
        \hline
    \end{tabular}}
    \caption{Comparison between the error on the amplitude $A$ for different benchmark cases, considering only the SKA auto-correlation and including also auto- and cross-correlations with astrometric probes.}
    \label{tab:ska_snr_table}
\end{table}

Finally, for a more quantitative understanding of the improvement in sensitivity generated by the inclusion of astrometric probes, we report in Tab.~\ref{tab:ska_snr_table} the Fisher matrix estimates of the error on the amplitude of the GWB for the benchmark scenarios discussed above.
The relative error~$\sigma_A/A$ on the GWB amplitude in the case of a pulsar-only measurement, for 5 years and 67 pulsars monitored, is approximately~$45\%$, while performing a joint-analysis reduces it to~$\sim 20\%$. 
In the best-case scenario, we obtain a relative error of~$16\%$ and~$13\%$, respectively.
We note that the difference between including or not astrometric correlations tends to decrease since the only parameters changing in this analysis are those related to timing residuals.
Thus, as the number of pulsars and the observation time increase, SKAO is able to detect higher multipoles that otherwise would not be seen, dominating the total SNR. 
However, an improvement in the angular sensitivity of astrometric observations would still be of immense value, not only in terms of sensitivity but also for accuracy, since it serves as a cross-check of the obtained constraints.


\section{Conclusions}
\label{sec:conclusions}

In this chapter, we investigate the enormous potential of SKAO to detect and characterize a GWB, with a particular focus on the synergy between PTA and astrometric measurements.
Our findings highlight the gain in taking a synergistic approach when looking at complementary GW probes because of the wealth of information contained in cross-correlations. 
Additionally, the analysis has been performed using the recent analytical results of~\cite{Perna:2025bdu}, which go beyond the common ``long-distance approximation''~\citep{Pyne:1995iy,Book:2010pf} and allow to take into account the spatial distribution of astronomical sources.

We compared the potential of SKAO alone and in combination with astrometric probes by performing a SNR analysis considering different SKA sensitivity curves corresponding to 5, 10, and 20 years of observations and a different number of pulsars. 
Our findings show that the cumulative SNR$^2$ increases by a factor of 5 already at $\ell=2$ in the most conservative scenario, where the observatory monitors 67 pulsars for 5 years. 
Moreover, our results suggest that, simply by monitoring the same number of pulsars of current PTA collaborations, SKAO is able to detect up to~$\ell = 4$ after only 5 years of observation time. 
On the other hand, in the optimistic scenario with up to~$10^3$ monitored pulsars over 20 years, we would detect up to~$\ell =10$.
This result could be further improved by astrometric observations if they achieve a higher angular precision.
We also performed a Fisher analysis to estimate the typical error bar on the amplitude of the GWB signal, obtaining an improvement of a factor 2 (from $\sigma_A/A\sim 45\%$ to $\sigma_A/A\sim 20\%$) when cross-correlations after 5 years and 67 pulsars monitored and up to $13\%$ in the optimal case of 1000 pulsars monitored for 20 years.

Additionally, in most cases of interest, the GWB can be considered stationary.
Therefore, cross-correlations between astrometric and timing residual datasets can be computed even for asynchronous experiments, i.e., if the two measurements are not taken at the same time. 
In other words, SKAO will provide an invaluable amount of \textit{legacy data} whose scientific potential could be exploited for decades after the experiment has ended.
In conclusion, our results suggest that SKAO will be pivotal for achieving a detection and characterization of the GWB, but also to unlock the constraining power that complementary measurements from other probes offer.


\section*{Acknowledgements}

DB \& GP acknowledge support from the COSMOS network (www.cosmosnet.it) through ASI (Italian Space Agency) Grants 2016-24-H.0, 2016-24-H.1-2018 and 2020-9-HH.0 GP aknowledges partial financial support from Fondazione Angelo Della Riccia and Fondazione Ing. Aldo Gini. GP thanks the Leibniz Universität in Hannover for the kind hospitality for the period during which this chapter was written.
NB acknowledges support from the European Union's Horizon Europe research and innovation program under the Marie Sk\l{}odowska-Curie grant agreement no.~101207487 (GWSKY - Mapping the Universe with Gravitational Waves) and by PRD/ARPE 2022 ``Cosmology with Gravitational waves and Large Scale Structure - CosmoGraLSS''.


\bibliographystyle{abbrvnat-maxbibnames4.bst}
\bibliography{bibliography} 

\end{document}